\documentclass[12pt,draftclsnofoot,onecolumn]{IEEEtran}
\usepackage{pifont}
\usepackage{graphicx}
\usepackage{fancyhdr}
\usepackage{amsmath,amssymb,amstext,amsfonts}
\usepackage{slashbox}
\usepackage{bm}
\usepackage{algorithm}
\usepackage{array}
\usepackage{cite}

\usepackage{subfigure}
\usepackage{balance}
\usepackage{algorithmicx}
\usepackage{algpseudocode}

\usepackage{braket}
\usepackage{tabulary}
\usepackage{diagbox}
\usepackage{amsthm}

\newtheorem{theorem}{Theorem}
\newtheorem{lemma}{Lemma}

\begin{document}

\title{Quantum Discrimination of Two Noisy Displaced Number States}

\author{Renzhi Yuan,~\IEEEmembership{Student Member,~IEEE} and Julian Cheng,~\IEEEmembership{Senior Member,~IEEE}.
\thanks{
Renzhi Yuan and Julian Cheng are with the School of Engineering, The University of British Columbia, Kelowna, BC, Canada (e-mails: renzhi.yuan@ubc.ca, julian.cheng@ubc.ca).
}
}

\maketitle

\begin{abstract}

The quantum discrimination of two non-coherent states draws much attention recently. In this letter, we first consider the quantum discrimination of two noiseless displaced number states. Then we derive the Fock representation of noisy displaced number states and address the problem of discriminating between two noisy displaced number states. We further prove that the optimal quantum discrimination of two noisy displaced number states can be achieved by the Kennedy receiver with threshold detection. Simulation results verify the theoretical derivations and show that the error probability of on-off keying modulation using a displaced number state is significantly less than that of on-off keying modulation using a coherent state with the same average energy.

\end{abstract}

\IEEEpeerreviewmaketitle

\section{Introduction}
% no \IEEEPARstart

Quantum discrimination of two optical quantum states plays a crucial rule in quantum information processing tasks, e.g., continuous-variable quantum key distributions and optical quantum communications \cite{silberhorn2002continuous,grosshans2003quantum,ghorai2019asymptotic,yuan2020free}. Due to the good compatibility with classical optical infrastructures, coherent states \cite{glauber1963coherent} generated by lasers are usually employed as the information carriers in quantum communication systems. However, the minimum discrimination error probability (MDEP) of discriminating two coherent states cannot be zero because of the non-orthogonal property of two coherent states \cite{glauber1963coherent}. To improve the performance of the quantum discrimination, new information carriers using non-coherent states draw much attention recently \cite{guerrini2019quantum,guerrini2020quantum}.

For example, the problem of discriminating between two noisy photon-added coherent states (PACSs) was addressed in \cite{guerrini2020quantum}. The PACS is generated by sequentially applying the displacement operator and the creation operator on a vacuum state. It was demonstrated that the error probability can be significantly reduced when PACSs instead of coherent states are employed in pulse position modulations \cite{guerrini2019quantum}. Inspired by \cite{guerrini2020quantum}, we focus on the quantum discrimination between two noisy displaced number states (DNSs). The DNS is generated by sequentially applying the creation operator and the displacement operator on a vacuum state. The properties of noiseless DNS were discussed in \cite{de1990properties,tanas1992phase,mo1996displaced}. However, the thermal noise is inevitable in preparing a DNS and the property of a noisy DNS has not been studied yet. Besides, to the best of the authors' knowledge, the problem of discriminating between two noisy DNSs has not been addressed.

In this letter, we first address the problem of discriminating between two noiseless DNSs. Then we derive the Fock representation of noisy DNSs and address the problem of discriminating between two noisy DNSs. Using the Fock representation of noisy DNSs, we further prove that the optimal quantum discrimination of two noisy DNSs can be achieved using a Kennedy receiver with a threshold detection. The simulation results verify our theoretical derivations. We also explore the possibility of employing DNSs instead of coherent states in on-off keying (OOK) modulations; and find that the error probability of OOK modulation using a DNS can be significantly reduced compared with the error probability of OOK modulation using a coherent state with the same average energy.

\section{Quantum Discrimination of Two Noiseless DNSs}\label{DiscriminateNoiselessDNS}
\subsection{Displaced Number State}
The DNS is generated by sequentially applying the creation operator and the displacement operator on a vacuum state; and it can be written as \cite{de1990properties,tanas1992phase}
\begin{equation}
\ket{\mu,k}= \hat{D}(\mu)\ket{k}
\end{equation}
\noindent where $\hat{D}(\mu)$ is the displacement operator; and $\ket{k}$ is the number state containing $k$ photons. The number state decomposition of a DNS can be obtained as \cite{tanas1992phase}
\begin{equation}
\ket{\mu,k}=\sum_{n=0}^{\infty}b_n\ket{n}
\end{equation}
\noindent where the coefficient $b_n$ is given by \cite{tanas1992phase}
\begin{equation}
b_n=
\left\{
\begin{array}{ll}
\sqrt{\frac{n!}{k!}}(-\mu^*)^{k-n}e^{-\frac{|\mu|^2}{2}}L_n^{(k-n)}(|\mu|^2), \text{ for } n<k\\
\sqrt{\frac{k!}{n!}}\mu^{n-k}e^{-\frac{|\mu|^2}{2}}L_n^{(n-k)}(|\mu|^2), \text{ for } n\geq k\\
\end{array}
\right.
\end{equation}
\noindent and where $L_n^{(a)}(x)$ is the generalized Laguerre polynomial of order $n$ with parameter $a$.

Using the number state decomposition, we can obtain the inner product of two DNSs $\ket{\mu,k}$ and $\ket{\xi,h}$, where we let $h\geq k$ without loss of generality, as
\begin{equation}\label{Superposition}
\begin{aligned}
\langle \xi,h|\mu,k \rangle=&\bra{h}\hat{D}(\mu-\xi)\ket{k}\\
=&\sum_{n=0}^{\infty}\bra{h}b_n(\mu-\xi,k)\ket{n}\\
=&\sqrt{\frac{k!}{h!}}(\mu-\xi)^{h-k}e^{-\frac{|\mu-\xi|^2}{2}}L_k^{(h-k)}(|\mu-\xi|^2)
\end{aligned}
\end{equation}
\noindent where in the first step we have used the properties of displacement operator: $\hat{D}(\alpha)=\hat{D}^{\dagger}(-\alpha)$ and $\hat{D}(\alpha)\hat{D}(\beta)=\hat{D}(\alpha+\beta)$.

\subsection{Discriminate Two Noiseless DNSs}
The key of discriminating between any two quantum states $\{\hat{\rho}_0, \hat{\rho}_1\}$ with prior probabilities $\{p_0, p_1\}$ is to find two positive operator-valued measure (POVM) operators $\{\hat{\Pi}_0,\hat{\Pi}_1\}$ that can minimize the discrimination error probability. According to the Helstrom's theory \cite{helstrom1969quantum}, the optimal POVM operators can be obtained as
$\hat{\Pi}_0=\sum_{\lambda_n<0}\ket{\lambda_n}\bra{\lambda_n}$ and $\hat{\Pi}_1=\hat{\mathbb{I}}-\hat{\Pi}_0$, where $\lambda_n$ and $\ket{\lambda_n}$ are the eigenvalue and the eigenvector of the decision operator $\hat{\Delta}=p_1\hat{\rho}_1-p_0\hat{\rho}_0$; $\hat{\mathbb{I}}$ is the identity operator. The MDEP of discriminating $\{\hat{\rho}_0, \hat{\rho}_1\}$ is obtained as the Helstrom bound \cite{helstrom1969quantum}
\begin{equation}\label{HelstromBound_1}
\begin{aligned}
P_e&=\frac{1}{2}(1-\|\hat{\Delta}\|_1)\\
&=p_1-\sum_{\lambda_n>0}\lambda_n
\end{aligned}
\end{equation}
\noindent where $\|\hat{A}\|_1=\text{tr}\{\sqrt{\hat{A}^{\dagger}\hat{A}}\}$ denotes the trace norm of the operator $\hat{A}$.

For discriminating between two pure states $\hat{\rho}_0=\ket{\psi_0}\bra{\psi_0}$ and $\hat{\rho}_1=\ket{\psi_1}\bra{\psi_1}$, the Helstrom bound \eqref{HelstromBound_1} can be rewritten as
\begin{equation}\label{HelstromBound_2}
P_e=\frac{1}{2}-\frac{1}{2}\sqrt{1-4p_0p_1|\left \langle \psi_0|\psi_1 \right \rangle|^2}.
\end{equation}

Therefore, the MDEP for discriminating between two noiseless DNSs $\ket{\xi,h}$ and $\ket{\mu,k}$ is determined by the inner product $\langle \xi,h|\mu,k \rangle$. Substituting \eqref{Superposition} into \eqref{HelstromBound_2}, we can obtain the MDEP as
\begin{equation}\label{P_e_noiseless}
\begin{aligned}
P_e=\frac{1}{2}-\frac{1}{2}&\left\{1-4p_0p_1\frac{k!}{h!}|\mu-\xi|^{2(h-k)}e^{-|\mu-\xi|^2}\right.\\
&\quad \times\left.\left[L_k^{(h-k)}(|\mu-\xi|^2)\right]^2\right\}^{\frac{1}{2}}.
\end{aligned}
\end{equation}

From \eqref{P_e_noiseless}, we can observe that the perfect discrimination with zero error probability happens in the following two situations: (i) $\mu=\xi$ and $h\neq k$; (ii) $L_k(|\mu-\xi|^2)=0$ and $h=k\neq 0$. Notice that when $h=k=0$, the two DNSs becomes two non-orthogonal coherent states $\ket{\xi}$ and $\ket{\mu}$. Then the MDEP approaches zero when $|\mu-\xi|^2$ approaches $\infty$.

\section{Quantum Discrimination of Two Noisy DNSs}
\subsection{Noisy DNSs}\label{NoisyDNS}
A noisy number state $\hat{\rho}(k)$ is obtained by applying creation operators on a thermal state $\hat{\rho}_{th}$, which results in
\begin{equation}
\hat{\rho}(k)=\frac{(\hat{A}^{\dagger})^k\hat{\rho}_{th}\hat{A}^k}{\text{tr}\{(\hat{A}^{\dagger})^k\hat{\rho}_{th}\hat{A}^k\}}
\end{equation}
\noindent where $\text{tr}\{\cdot\}$ denotes the trace operation.

Then the noisy DNS $\hat{\rho}(\mu,k)$ is defined as
\begin{equation}\label{noisy DNS}
\hat{\rho}(\mu,k)\triangleq\frac{\hat{D}(\mu)(\hat{A}^{\dagger})^k\hat{\rho}_{th}\hat{A}^k\hat{D}^{\dagger}(\mu)}{N_k}
\end{equation}
\noindent where $N_k$ can be obtained as
\begin{equation}
\begin{aligned}
N_k=\text{tr}\{\hat{D}(\mu)(\hat{A}^{\dagger})^k\hat{\rho}_{th}\hat{A}^k\hat{D}^{\dagger}(\mu)\}=k!(n_t+1)^k
\end{aligned}
\end{equation}
\noindent and where $n_t$ is the average number of thermal photons due to the presence of thermal noise. The following theorem presents the Fock representation for a noisy DNS.

\begin{theorem}{(Fock representation)}\label{theorem_Fock_representation}
The Fock representation of a noisy DNS $\hat{\rho}(\mu,k)$ is found to be
\begin{equation}
\begin{aligned}
\label{Fock representation}
\bra{n}&\hat{\rho}(\mu,k)\ket{m}\\
&=\sum_{i=0}^{k}\sum_{j=0}^{k}I(n\geq i;m\geq j)\frac{(-1)^{i+j}\binom{m}{j}\binom{k}{i}}{(k-j)!}\sqrt{\frac{n!}{m!}}e^{-\frac{|\mu|^2}{n_t+1}}\\
&\quad \times \frac{|\mu|^{2(k-j)}(\mu^*)^{m-n} n_t^{n-i}}{(n_t+1)^{m+k-j+1}}L_{n-i}^{(m-n+i-j)}\left(-\frac{|\mu|^2}{n_t(n_t+1)}\right)
\end{aligned}
\end{equation}
\noindent where $I(n\geq i;m\geq j)$ is an indicator function defined as
\begin{equation}
\begin{aligned}
I(n\geq i;m\geq j)\triangleq\left\{
\begin{array}{ll}
1,\text{ for } n\geq i \text{ and } m\geq j \\
0,\text{ otherwise.}
\end{array}\right.
\end{aligned}
\end{equation}
\end{theorem}

\begin{proof}
See Appendix \ref{A1}.
\end{proof}

Using the Fock representation in \eqref{Fock representation}, we can obtain the photon statistics of a noisy DNS as
\begin{equation}
\begin{aligned}
p(n)&= \bra{n}\hat{\rho}(\mu,k)\ket{n}\\
&=\sum_{i=0}^{k}\sum_{j=0}^{k}I(n\geq i;n\geq j)\frac{(-1)^{i+j}\binom{n}{j}\binom{k}{i}}{(k-j)!}e^{-\frac{|\mu|^2}{n_t+1}}\\
&\quad \times \frac{|\mu|^{2(k-j)} n_t^{n-i}}{(n_t+1)^{n+k-j+1}}L_{n-i}^{(i-j)}\left(-\frac{|\mu|^2}{n_t(n_t+1)}\right).
\end{aligned}
\end{equation}

Then the average number of photons $n_p$ contained in a noisy DNS can be obtained in the following lemma.

\begin{lemma}\label{Lemma_1}
The average number of photons $n_p(\mu,k)$ contained in a noisy DNS $\hat{\rho}(\mu,k)$ is found to be
\begin{equation}\label{n_p}
n_p(\mu,k)=|\mu|^2+k(n_t+1)+n_t.
\end{equation}
\end{lemma}
\begin{proof}
See Appendix \ref{A2}.
\end{proof}

\subsection{Discriminate Two Noisy DNSs}\label{DiscriminateNoisyDNS}
Now we consider the discrimination of two noisy DNSs $\hat{\rho}_0=\hat{\rho}(\xi,h)$ and $\hat{\rho}_1=\hat{\rho}(\mu,k)$ with prior probabilities $p_0$ and $p_1$, respectively. Here we let $h\geq k$ without loss of generality. According to the Helstrom bound \eqref{HelstromBound_1}, the MDEP is determined by all the positive eigenvalues $\lambda_n$ of the decision operator
$\hat{\Delta}=p_1\hat{\rho}(\mu,k)-p_0\hat{\rho}(\xi,h)$. Because the displacement operator is an unitary operator, $\hat{\Delta}$ and $\hat{D}(-\xi)\hat{\Delta}\hat{D}^{\dagger}(-\xi)$ share the same eigenvalues. Then it is readable to show that
\begin{equation}\label{tracenorm_2}
\|p_1\hat{\rho}(\mu,k)-p_0\hat{\rho}(\xi,h)\|_1=\|p_1\hat{\rho}(\mu-\xi,k)-p_0\hat{\rho}(0,h)\|_1.
\end{equation}

It is challenging to obtain an analytical form of the eigenvalues $\lambda_n$ for an arbitrary $\hat{\Delta}$. However, eq. \eqref{tracenorm_2} indicates that when $\mu=\xi$, we only need to obtain the eigenvalues of $p_1\hat{\rho}(0,k)-p_0\hat{\rho}(0,h)$. Using the Fock representation \eqref{Fock representation}, we can obtain the eigenvalues of $p_1\hat{\rho}(0,k)-p_0\hat{\rho}(0,h)$ as
\begin{equation}\label{Lambda_n}
\begin{aligned}
\lambda_n&=p_1\bra{n}\hat{\rho}(0,k)\ket{n}-p_0\bra{n}\hat{\rho}(0,h)\ket{n}\\
&=
\left\{
\begin{array}{ll}
0, \text{ for } n<k\\
p_1\binom{n}{k}\frac{n_t^{n-k}}{(n_t+1)^{n+1}}, \text{ for } k\leq n<h\\
p_1\binom{n}{k}\frac{n_t^{n-k}}{(n_t+1)^{n+1}}-p_0\binom{n}{h}\frac{n_t^{n-h}}{(n_t+1)^{n+1}}, \text{ for } n\geq h.
\end{array}
\right.
\end{aligned}
\end{equation}

Now the key of obtaining a tractable MDEP is to find all the positive eigenvalues. To achieve this, in the following we first introduce the Kennedy receiver with threshold detection \cite{yuan2020kennedy}, and then we prove that the Kennedy receiver with threshold detection can achieve the optimal discrimination and provide a tractable MDEP.

\subsection{Kennedy Receiver with Threshold Detection}\label{KennedyDetection}

A Kennedy receiver with threshold detection \cite{yuan2020kennedy} consists of a displacement operator $\hat{D}(\beta)$ and a photon counting process followed by a threshold detection based on the counted photons. The threshold detection is characterized by two POVM operators \begin{equation}\label{ThreholdOperators}
\hat{M}_0=\hat{\mathbb{I}}-\sum_{n=0}^{n_{th}}\ket{n}\bra{n}; \quad \quad \hat{M}_1=\sum_{n=0}^{n_{th}}\ket{n}\bra{n}
\end{equation}
\noindent where $n_{th}$ is the detection threshold of the counting photons. These two POVM operators correspond to the following threshold detection rule
\begin{equation}
\label{Threshold test}
n \mathop{\lesseqgtr} \limits_{\hat{\rho}_0}^{\hat{\rho}_1} n_{th}.
\end{equation}

If we set the displacement operator as $\hat{D}(\beta)=\hat{D}(-\mu)$, then the input states $\hat{\rho}(\mu,h)$ and $\hat{\rho}(\mu,k)$ are displaced as $\hat{\rho}(0,h)$ and $\hat{\rho}(0,k)$, respectively. Then the error probability of the receiver can be calculated by
\begin{equation}\label{P_e_2}
\begin{aligned}
P_e&=p_0\text{tr}\{\hat{M}_1\hat{\rho}(0,h)\}+p_1\text{tr}\{\hat{M}_0\hat{\rho}(0,k)\}.
\end{aligned}
\end{equation}
\noindent Substituting \eqref{ThreholdOperators} into \eqref{P_e_2}, we can obtain
\begin{equation}\label{P_e_3}
\begin{aligned}
P_e&=p_1-\sum_{n=0}^{n_{th}}\left(p_1\bra{n}\hat{\rho}(0,k)\ket{n}-p_0\bra{n}\hat{\rho}(0,h)\ket{n}\right)\\
&=p_1-\sum_{n=0}^{n_{th}}\lambda_n.
\end{aligned}
\end{equation}

The optimal threshold $n_{th}$ is obtained by minimizing the error probability in \eqref{P_e_3}. The following optimal discrimination theorem guarantees that the Kennedy receiver with optimal threshold $n_{th}$ can always achieve the MDEP.

\begin{theorem}{(Optimal discrimination)}\label{Theorem_optimal_discrimination}
The optimal discrimination of two noisy DNSs $\hat{\rho}(\mu,h)$ and $\hat{\rho}(\mu,k)$ can be achieved by the Kennedy receiver with threshold detection, where the displacement operator is $\hat{D}(-\mu)$; and the MDEP can be obtained as
\begin{equation}\label{P_e}
P_e=p_1-\sum_{n=0}^{n_{th}}\lambda_n
\end{equation}
\noindent where the optimal threshold $n_{th}$ is the maximum $n$ satisfying
\begin{equation}\label{optimalthreshold}
\begin{aligned}
\binom{n}{k}n_t^{h-k}\geq \binom{n}{h}\frac{p_0}{p_1}.
\end{aligned}
\end{equation}
\end{theorem}

\begin{proof}
See Appendix \ref{A3}.
\end{proof}

Although the POVM operators of optimal quantum discrimination for two quantum states can be obtained from the Helstrom's theory, the realization of the optical quantum discrimination is usually intractable. However, Theorem \ref{Theorem_optimal_discrimination} indicates that the optimal quantum discrimination of two noisy DNSs $\hat{\rho}(\mu,h)$ and $\hat{\rho}(\mu,k)$ is realizable by the Kennedy receiver with threshold detection \footnote{Note that the Kennedy receiver with threshold detection is a near-optimum receiver for discriminating between two coherent states.}.

\section{Numerical Results}
\label{sect:NumericalResults}
\begin{figure}
\begin{center}
\includegraphics[width=0.48\textwidth, draft=false]{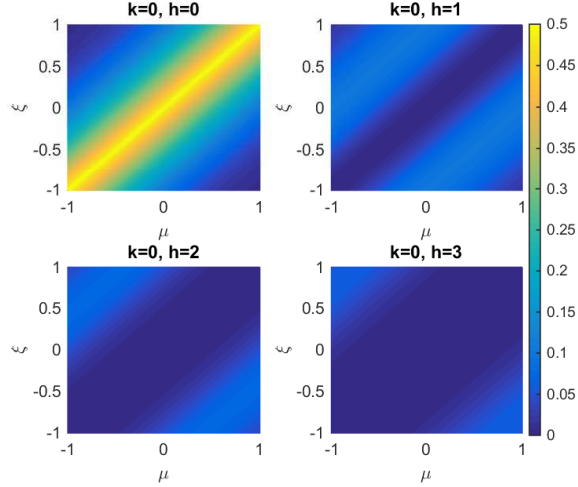}
\caption{MDEP for discriminating two noiseless DNSs ($n_t=0$)}
\vspace{-0.4cm}
\label{Fig:TwoNoiselessDNS}
\end{center}
\end{figure}

\begin{figure}
\begin{center}
\includegraphics[width=0.48\textwidth, draft=false]{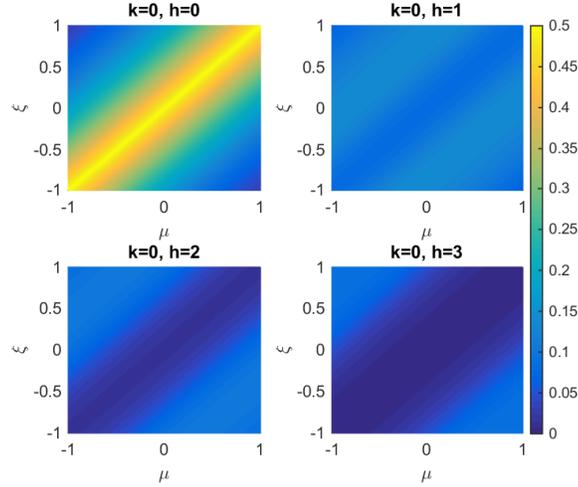}
\caption{MDEP for discriminating two noisy DNSs ($n_t=0.2$)}
\vspace{-0.4cm}
\label{Fig:TwoNoisyDNS}
\end{center}
\end{figure}

The prior probabilities are set as $p_0=p_1=0.5$ in this section. Figs. \ref{Fig:TwoNoiselessDNS} and \ref{Fig:TwoNoisyDNS} present the MDEP for discriminating two noiseless DNSs and two noisy DNSs, respectively. From Fig. \ref{Fig:TwoNoiselessDNS}, we can observe that when $k=h$, the MDEP decreases as $|\mu-\xi|$ increases; and when $k\neq h$, the MDEP achieves zero when $\mu=\xi$. Besides, when $k\neq h$, the MDEP decreases as the gap $h-k$ increases. Comparing Fig. \ref{Fig:TwoNoisyDNS} with Fig. \ref{Fig:TwoNoiselessDNS}, we can see that the MDEP for discriminating two noisy DNSs demonstrates similar properties to that for discriminating two noiseless DNSs. Besides, the MDEP for discriminating two noisy DNSs is always larger than that for discriminating two noiseless DNSs with the same parameters.

\begin{figure}
\begin{center}
\includegraphics[width=0.48\textwidth, draft=false]{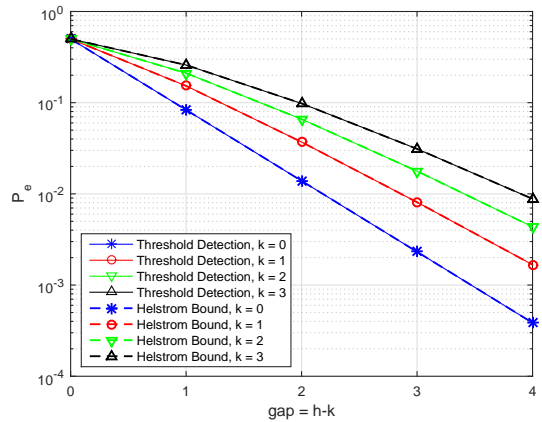}
\caption{Error probabilities under different gap ($\mu=\xi=1, n_t=0.2$)}
\vspace{-0.4cm}
\label{Fig:P_e_vs_gap}
\end{center}
\end{figure}

Next we check the error probability for discriminating between two noisy DNSs obtained by the Kennedy receiver with threshold detection with $\mu=\xi=1$ under different gaps, shown in Fig. \ref{Fig:P_e_vs_gap}. We also plot the MDEP results of Helstrom bound obtained by the optimal quantum discrimination. We can see that the error probabilities obtained by the Kennedy receiver with threshold detection coincide with the MDEP results of Helstrom bound obtained by the optimal quantum discrimination. This verifies the result of Theorem \ref{Theorem_optimal_discrimination}. The MDEP decreases as the gap increases for a given $k$, which is as expected. Besides, when the gap is fixed, the MDEP decreases as $k$ decreases. For a given gap, a smaller $k$ implies a smaller energy requirement. This indicates that if we use DNSs as the information carriers in intensity modulations, a smaller $k$ can achieve a better performance in not only error probability but also energy efficiency. Therefore, the OOK modulation with $k=0$ for a given energy gap is the optimal intensity modulation in terms of both the error probability and the energy efficiency.

\begin{figure}
\begin{center}
\includegraphics[width=0.48\textwidth, draft=false]{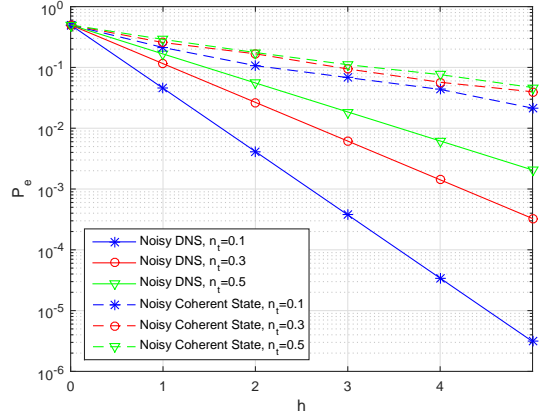}
\caption{Error probabilities of OOK modulation under different thermal noises}
\vspace{-0.4cm}
\label{Fig:DNS_vs_Coherent}
\end{center}
\end{figure}

At last, we consider a special case of discriminating two noisy DNSs with $\mu=\xi=0$ and $k=0$ under different gaps $h$, which corresponds to an OOK modulation in communication systems. The error probabilities under different $h$ with different thermal noises are shown in Fig. \ref{Fig:DNS_vs_Coherent}. We also plot the error probabilities of the OOK modulation employing a coherent state with the same average energy per information bit. We can see that the error probability decreases as $h$ increases, which is as expected. Besides, we can also see that the error probability of OOK modulations employing a DNS can be significantly reduced compared with that of employing a coherent state with the same average energy.

\section{Conclusion}
\label{sect:Conclusion}

We addressed the problem of discriminating between two noisy DNSs. We first considered the quantum discrimination of two noiseless DNSs. Then we derived the Fock representation of a noisy DNS, and then used the Fock representation to derive the MDEP of discriminating two noisy DNSs. We further proved that the optimal quantum discrimination of two noisy DNSs can be achieved by the Kennedy receiver with threshold detection. The simulation results verified our theoretical derivations. Besides, we found that the error probability of OOK modulation employing a DNS is significantly less than that of OOK modulation employing a coherent state with the same average energy.

\bibliographystyle{IEEEtran}
\bibliography{refabrv}

\appendices
\section{Proof of Theorem 1}\label{A1}
We use the coherent-state representation of noisy DNS $\hat{\rho}(\mu,k)$ to obtain its Fock representation. The coherent-state representation of a noisy DNS can be obtained by \cite{glauber1963coherent}
\begin{equation}\label{R_function_1}
\begin{aligned}
R(\alpha^*,\beta)&=e^{\frac{1}{2}|\alpha|^2+\frac{1}{2}|\beta|^2}\bra{\alpha}\hat{\rho}(\mu,k)\ket{\beta}.
\end{aligned}
\end{equation}
\noindent Substituting \eqref{noisy DNS} into \eqref{R_function_1}, we obtain
\begin{equation}
\begin{aligned}
R(\alpha^*,\beta)&=\frac{e^{\frac{1}{2}|\alpha|^2+\frac{1}{2}|\beta|^2}}{k!(n_t+1)^k}\bra{\alpha-\mu}(\hat{A}^{\dagger})^k\hat{\rho}_{th}\hat{A}^k\ket{\beta-\mu}\\
&=\frac{e^{\frac{1}{2}|\alpha|^2+\frac{1}{2}|\beta|^2}(\alpha^*-\mu^*)^k(\beta-\mu)^k}{k!(n_t+1)^k}\bra{\alpha}\hat{\rho}_{th}(\mu)\ket{\beta}
\end{aligned}
\end{equation}
\noindent where $\hat{\rho}_{th}(\mu)$ is the displaced thermal state. Using the coherent-state representation $R_{th}(\alpha^*,\beta)$ of displaced thermal state \cite{guerrini2020quantum}, we can obtain
\begin{equation}
\begin{aligned}
R(\alpha^*&,\beta)\\
&=\frac{(\alpha^*-\mu^*)^k(\beta-\mu)^k}{k!(n_t+1)^k}R_{th}(\alpha^*,\beta)\\
&=\sum_{n=0}^{\infty}\sum_{m=0}^{\infty}\frac{(\alpha^*)^n\beta^m}{\sqrt{n!m!}}\sum_{i=0}^{k}\sum_{j=0}^{k}I(n\geq i;m\geq j)\frac{(-1)^{i+j}}{(k-j)!}\\
&\quad \times \binom{m}{j}\binom{k}{i}\sqrt{\frac{n!}{m!}}e^{-\frac{|\mu|^2}{n_t+1}}|\mu|^{2(k-j)}(\mu^*)^{m-n}\\
&\quad \times \frac{n_t^{n-i}}{(n_t+1)^{m+k-j+1}}L_{n-i}^{(m-n+i-j)}\left(-\frac{|\mu|^2}{n_t(n_t+1)}\right).
\end{aligned}
\end{equation}
\noindent Using the relation between the coherent-state representation and the Fock representation \cite{glauber1963coherent}, we obtain the Fock representation of noisy DNS as \eqref{Fock representation}.

\section{Proof of Lemma 1}\label{A2}
The displacement operator is defined as
\begin{equation}
\begin{aligned}
\hat{D}(\alpha)\triangleq e^{\alpha \hat{A}^{\dagger}-\alpha^*\hat{A}}=e^{-\frac{1}{2}|\alpha|^2}e^{\alpha\hat{A}^{\dagger}}e^{-\alpha^*\hat{A}}.
\end{aligned}
\end{equation}
\noindent Using the Taylor series of matrix exponential $e^{\alpha\hat{A}^{\dagger}}=\sum_{k=0}^{\infty}\frac{\alpha^k}{k!}(\hat{A}^{\dagger})^k$ and the commutator $[\hat{A}^{\dagger},\hat{A}]=\hat{\mathbb{I}}$, we can obtain the following commutators
\begin{equation}\label{Commutator_1}
[\hat{A},\hat{D}(\alpha)]=\alpha\hat{D}(\alpha);\quad [\hat{A}^{\dagger},\hat{D}(\alpha)]=\alpha^*\hat{D}(\alpha).
\end{equation}
\noindent The average number of photons $n_p(\mu,k)$ of a noisy DNS $\hat{\rho}(\mu,k)$ is defined as
\begin{equation}\label{n_p_1}
\begin{aligned}
n_p(\mu,k)&\triangleq\text{tr}\{\hat{\rho}(\mu,k)\hat{A}^{\dagger}\hat{A}\}\\
&=\text{tr}\{\hat{\rho}(0,k)\hat{D}^{\dagger}(\mu)\hat{A}^{\dagger}\hat{A}\hat{D}(\mu)\}.
\end{aligned}
\end{equation}
\noindent Using the commutators in \eqref{Commutator_1}, we can obtain
\begin{equation}\label{temp_1}
\hat{D}^{\dagger}(\mu)\hat{A}^{\dagger}\hat{A}\hat{D}(\mu)=(\hat{A}^{\dagger}\hat{A}+\mu\hat{A}^{\dagger}+\mu^*\hat{A}+|\mu|^2\hat{\mathbb{I}}).
\end{equation}
\noindent Substituting \eqref{temp_1} into \eqref{n_p_1}, we can obtain
\begin{equation}\label{n_p_2}
\begin{aligned}
n_p(\mu,k)&=|\mu|^2+\text{tr}\{\hat{\rho}(0,k)\hat{A}^{\dagger}\hat{A}\}+\mu\text{tr}\{\hat{\rho}(0,k)\hat{A}^{\dagger}\}\\
&\quad +\mu^*\text{tr}\{\hat{\rho}(0,k)\hat{A}\}
\end{aligned}
\end{equation}
\noindent where $\text{tr}\{\hat{\rho}(0,k)\hat{A}^{\dagger}\hat{A}\}=k(n_t+1)+n_t$ is the average number of photons of a photon-added thermal state \cite[eq. (19)]{guerrini2020quantum}. Note that
\begin{equation}
\text{tr}\{\hat{\rho}(0,k)\hat{A}^{\dagger}\}=\text{tr}\{\hat{\rho}(0,k)\hat{A}\}=0
\end{equation}
\noindent where we have used the property that $P(\alpha)|_{\mu=0}$ is an even function of $\alpha$. Therefore, we have $n_p(\mu,k)=|\mu|^2+k(n_t+1)+n_t$.

\section{Proof of Theorem 2}\label{A3}
From \eqref{Lambda_n}, we can observe that $\lambda_n\geq 0$ for any $n<h$. Then the key is to find all positive $\lambda_n$ when $n\geq h$. Note that if $\lambda_n<0$ when $n\geq h$, then we have
\begin{equation}\label{inequality_1}
p_1\binom{n}{k}n_t^{h-k}<p_0\binom{n}{h}.
\end{equation}
Then for $\lambda_{n+1}$, we have
\begin{equation}\label{inequality_2}
\begin{aligned}
\lambda_{n+1}&=\frac{n_t^{n+1-h}}{(n_t+1)^{n+2}}\left[\frac{n+1}{n+1-k}p_1\binom{n}{k}n_t^{h-k}\right.\\
&\quad \quad \quad\quad \quad \quad\quad \quad \quad \left.-\frac{n+1}{n+1-h}p_0\binom{n}{h}\right]\\
&\leq \frac{n_t^{n+1-h}}{(n_t+1)^{n+2}}\frac{n+1}{n+1-h}\left[p_1\binom{n}{k}n_t^{h-k}-p_0\binom{n}{h}\right]
\end{aligned}
\end{equation}
\noindent where we have used the inequality $\frac{n+1}{n+1-k}\leq \frac{n+1}{n+1-h}$ for $h\geq k$. According to \eqref{inequality_1}, we have $\lambda_{n+1}<0$. In other words, $\lambda_{n}<0$ guarantees $\lambda_{n+1}<0$. This indicates that there exists a threshold $n_{th}$ such that
\begin{equation}
\left\{
\begin{array}{ll}
\lambda_n\geq 0, \text{ } n\leq n_{th}\\
\lambda_n < 0,  \text{ } n>n_{th}.
\end{array}
\right.
\end{equation}

Therefore, by optimizing the threshold in \eqref{P_e_3}, we can always achieve the MDEP in \eqref{HelstromBound_1}, i.e., we have
\begin{equation}
\sum_{\lambda_n>0}\lambda_n=\sum_{n=0}^{n_{th}}\lambda_n.
\end{equation}
\noindent Accordingly, the optimal threshold $n_{th}$ is the maximum $n$ satisfying \eqref{optimalthreshold}.

\balance
\end{document}